\documentclass[12pt]{article}

\usepackage[T1]{fontenc}
\usepackage[utf8]{inputenc}
\usepackage{lmodern}
\usepackage{microtype}
\usepackage[margin=1in]{geometry}
\usepackage{amsmath,amssymb}
\usepackage{graphicx}
\usepackage{setspace}
\usepackage{booktabs,longtable,array,tabularx, multirow, rotating}
\usepackage{float}
\usepackage{caption}
\usepackage{subcaption}
\usepackage[dvipsnames]{xcolor}
\usepackage[round,authoryear]{natbib}
\usepackage[colorlinks=true,linkcolor=blue,citecolor=blue,urlcolor=blue]{hyperref}

\graphicspath{{figures/}}
\definecolor{revision}{RGB}{180,30,30}

\title{\textbf{Learning Volatility Dependence Networks in UK Equity Markets using Penalised Spatiotemporal ARCH Models}}

\author{Elkanah Nyabuto$^{1}$\thanks{$^{*}$Corresponding author: \texttt{e.nyabuto.1@research.gla.ac.uk}} \and Philipp Otto$^{1}$}

\date{\small $^{1}$School of Mathematics and Statistics, University of Glasgow}

\begin{document}

\maketitle

\begin{abstract}
Spatiotemporal ARCH models capture temporal volatility persistence and cross-sectional dependence but typically require a predefined spatial weight matrix. This is restrictive in financial markets, where the dependence network is rarely known. We develop a LASSO-penalised quasi-maximum likelihood estimator that jointly learns a sparse weight matrix and estimates temporal dependence and covariate effects. Monte Carlo experiments show that the method recovers the model parameters and underlying network, with accuracy improving as the temporal sample size increases. We apply the method to daily returns from twenty UK-listed firms and compare the learned network with Euclidean-distance, correlation, autoregressive-similarity and sector-based structures. The learned network improves out-of-sample volatility prediction and reveals directional firm-level and cross-sector dependence not captured by the predefined alternatives. Glencore emerges as the strongest source of conditional-volatility dependence, while temporal persistence is concentrated among fewer than half of the firms. The method provides a data-driven framework for learning interpretable conditional-volatility networks.
\end{abstract}

\noindent\textbf{Keywords:} Conditional volatility; financial networks; LASSO; spatial weight matrix; spatiotemporal ARCH; Volatility spillovers.

\newpage
\setstretch{1} 

\section{Introduction}
Financial markets are inherently interconnected systems in which shocks originating in one asset, sector, or market segment can propagate rapidly throughout the broader financial network. These interactions give rise to complex dependence structures characterised by volatility clustering, cross-sectional spillovers, and time-varying correlations among assets \citep{engle1982autoregressive,bollerslev1986generalized,engle2002dynamic}. Volatility clustering, whereby periods of elevated volatility tend to be followed by further periods of elevated volatility, reflects strong temporal persistence in financial returns. At the same time, economic linkages, common information channels, and investor behaviour generate cross-sectional dependencies that facilitate the transmission of volatility across firms and markets \citep{diebold2012better,diebold2014network}. Understanding these interconnected dynamics is fundamental to modern financial econometrics, with important implications for risk management, portfolio allocation, asset pricing, and the monitoring of systemic risk.

Classical volatility models such as the Autoregressive Conditional Heteroskedasticity (ARCH) model of \citet{engle1982autoregressive} and the Generalised ARCH (GARCH) model of \citet{bollerslev1986generalized} provide powerful tools for modelling temporal dependence in conditional variances. These frameworks have been widely applied in financial economics and continue to form the foundation of modern volatility modelling. However, their standard formulations treat individual volatility processes independently and therefore fail to explicitly account for contemporaneous interactions between assets. In practice, volatility spillovers are pervasive, particularly during periods of financial distress when shocks may spread rapidly across institutions, sectors, or geographical markets \citep{forbes2002no,diebold2009measuring}. Ignoring such interdependencies may result in incomplete representations of risk transmission mechanisms and lead to suboptimal forecasting performance. To address these limitations, multivariate volatility models have been developed to capture dynamic relationships among multiple financial series. Notable examples include the BEKK model \citep{engle1995multivariate}, the Dynamic Conditional Correlation (DCC) model \citep{engle2002dynamic}, and the Generalised Orthogonal GARCH framework \citep{van2002go}. While these models accommodate time-varying covariance structures, their parameterisation becomes increasingly complex as the number of assets grows. Consequently, estimation may become computationally demanding and difficult to interpret in high-dimensional settings \citep{bauwens2006multivariate}. Moreover, these models do not directly identify the pathways through which shocks and volatility propagate across assets contemporaneously. This limitation has motivated the development of network and spatial modelling approaches that explicitly represent dependence structures between units.

In these frameworks, cross-sectional interactions are represented through a dependence matrix describing the strength and direction of relationships between observational units. Building on classical volatility models, Sato and Matsuda \citep{sato2017spatial} introduced the Spatial ARCH (S-ARCH) model, which was subsequently extended to the Spatial GARCH (S-GARCH) framework \citep{sato2021spatial}. These spatial volatility models were further developed in spatiotemporal settings by \citet{otto2018generalised,otto2024dynamic,otto2024multivariate}. Together, these approaches provide a flexible framework for representing volatility dependence through explicitly specified spatial or network structures \citep[see][for a review]{otto2025spatial}. However, they generally assume that the spatial weighting matrix is known a priori and specified using application-specific information such as geographical proximity, sectoral relationships, or other externally defined notions of connectivity. While these specifications may be appropriate in some settings, they are often difficult to justify in financial systems, where volatility transmission is driven by complex and evolving economic interactions that are not directly observable. Consequently, misspecification of the spatial weights matrix may distort the estimated volatility network, bias parameter estimates, and lead to misleading conclusions regarding volatility spillovers and systemic risk. The underlying spatiotemporal conditional-variance specification is therefore not itself the methodological contribution of this paper. Rather, the contribution concerns estimation of the contemporaneous dependence structure when the spatial weight matrix is unknown, instead of specifying this structure externally before model estimation.

Relaxing the assumption of a predefined dependence structure has become an active area of research within spatial econometrics. Several authors have proposed methodologies for estimating interaction matrices directly from observed data rather than imposing them a priori. Early contributions by \citet{bhattacharjee2006estimation} and \citet{bhattacharjee2013estimation} considered the estimation of spatial weights matrices under structural constraints. Subsequently, \citet{ahrens2015two} developed a two-step LASSO-based procedure for recovering sparse spatial weight matrices, while \citet{lam2020estimation} proposed a penalised estimation framework for spatial lag models in high-dimensional settings. More recently, \citet{zhang2018spatial} investigated model averaging and spatial weights matrix selection, and \citet{merk2022estimation} demonstrated the effectiveness of adaptive LASSO techniques for recovering spatial dependence structures. Collectively, these studies demonstrate a growing shift towards data-driven estimation of spatial dependence structures and illustrate the advantages of sparse regularisation for recovering interpretable interaction networks in high-dimensional settings. Nevertheless, most of this literature has focused on conditional mean models, with comparatively little attention devoted to estimating unknown dependence structures within spatiotemporal volatility models. Extending data-driven dependence estimation to volatility models is non-trivial because dependence enters through the conditional variance rather than the conditional mean. This distinction motivates estimation procedures capable of recovering volatility dependence structures directly from observed data.

Sparse regularisation provides a natural framework for this purpose because it simultaneously performs parameter estimation and dependence selection while controlling model complexity. In particular, the Least Absolute Shrinkage and Selection Operator (LASSO) has demonstrated considerable success in recovering sparse dependence structures in high-dimensional statistical models \citep{tibshirani1996regression,otto2023estimation}. By shrinking weak interactions towards zero while retaining dominant dependence relationships, LASSO produces interpretable network structures, reduces overfitting, and improves estimation stability. These properties are especially attractive in financial applications, where the number of potential cross-sectional interactions increases rapidly with the number of assets. The methodological contribution considered here does not arise from the LASSO penalty itself, which is well established, but from its incorporation into the joint estimation problem generated by an unknown contemporaneous interaction matrix in a conditional-variance model. In this setting, the dependence coefficients must be estimated simultaneously with temporal persistence and covariate effects while satisfying the structural restrictions required for a well-defined spatial multiplier.

Motivated by these challenges, this paper develops a penalised estimation framework for learning an unknown contemporaneous dependence matrix within an existing class of spatiotemporal conditional-variance models. The contribution therefore lies not in introducing a new Spatiotemporal ARCH specification, but in treating the dependence matrix as an unknown model component and estimating it jointly with temporal persistence and covariate effects under appropriate structural constraints. Specifically, we formulate a LASSO-penalised quasi-maximum likelihood estimator that performs simultaneous parameter estimation and sparse dependence-network selection. The contribution is therefore an integration of sparse dependence learning with a constrained spatiotemporal volatility estimation problem in which the network enters the conditional variance rather than the conditional mean. The methodological question is whether the latent contemporaneous dependence structure can be learned jointly with the remaining volatility parameters and whether doing so provides empirical advantages over plausible externally specified networks. The methodology is evaluated through an extensive Monte Carlo simulation study and illustrated using financial market data to demonstrate its ability to recover latent volatility transmission networks in high-dimensional settings. More broadly, the proposed framework extends sparse dependence learning from conditional mean models to conditional variance models, thereby contributing to the growing literature on data-driven spatiotemporal modelling.

The paper makes three specific contributions. First, it extends data-driven estimation of unknown spatial dependence structures from conditional-mean models to a spatiotemporal conditional-variance setting. Second, it formulates a constrained LASSO-penalised quasi-likelihood procedure for jointly estimating the contemporaneous dependence matrix, temporal persistence parameters, and covariate effects. Third, it evaluates the resulting estimator through finite-sample experiments and an empirical comparison with plausible predefined dependence structures, allowing the benefits of learning the network to be distinguished from those arising merely from imposing sparsity. The remainder of the paper is organised as follows. Section \ref{sec:ARCH_Model} introduces the Sp-ARCH model and the underlying dependence structure. Section \ref{sec:estimation} presents the penalised estimation methodology and regularisation framework. Section \ref{sec:simulationstudy} reports the results of the simulation study. Section \ref{sec:stock_application} applies the proposed methodology to financial market data and discusses the resulting volatility network. Section \ref{sec:conclusion} concludes and outlines directions for future research.

\section{Spatiotemporal ARCH Model}\label{sec:ARCH_Model}
Consider a spatiotemporal panel process drawn from a univariate random field $\{Y_t(\boldsymbol{s}): t=1,\ldots, T;\,\boldsymbol{s}\in D_S\}$, where $D_S=\{\boldsymbol{s}_1,\boldsymbol{s}_2,\ldots,\boldsymbol{s}_n\}$ denotes a finite set of $n$ spatial units. In financial applications, these units may represent firms, sectors, or other interconnected economic entities, while $t$ indexes discrete time points. The term ``spatial'' is retained to reflect the spatial-autoregressive formulation of the model and its relationship to the Spatial and Spatiotemporal ARCH literature. More generally, however, the index $\boldsymbol{s}_i$ identifies a cross-sectional unit and need not correspond to a geographical location. Let $\boldsymbol{Y}_t=(Y_t(\boldsymbol{s}_1),\ldots,Y_t(\boldsymbol{s}_n))^\top$ denote the vector of observations across all units at time $t$. Then, the process is assumed to follow the conditionally heteroskedastic structure
\begin{equation}
    \boldsymbol{Y}_t=\boldsymbol{h}_t^{1/2}\boldsymbol{\varepsilon}_t,
    \label{eq: spmodel}
\end{equation}
where $\boldsymbol{h}_t=(h_t(\boldsymbol{s}_1),\ldots,h_t(\boldsymbol{s}_n))^\top$ denotes the vector of conditional variances, such that $\sqrt{h_t(\boldsymbol{s}_i)}$ represents the conditional volatility of unit $i$ at time $t$, and $\boldsymbol{\varepsilon}_t=(\varepsilon_t(\boldsymbol{s}_1),\ldots,\varepsilon_t(\boldsymbol{s}_n))^\top$ is a vector of innovations satisfying $\mathbb{E}(\boldsymbol{\varepsilon}_t)=\mathbf{0}$ and $\mathbb{E}(\boldsymbol{\varepsilon}_t\boldsymbol{\varepsilon}_t^\top)=\mathbf{I}_n$. The log conditional variance dynamics are specified as
\begin{equation}
\log \boldsymbol{h}_t=\mathbf{W}\log(\boldsymbol{Y}_t^2)+\sum_{p=1}^{P}\mathbf{\Phi}_{p}\log(\boldsymbol{Y}_{t-p}^2)
+\mathbf{X}_t\boldsymbol{\beta}+\boldsymbol{\mu},\label{eq:log volatility}
\end{equation}
where $\log(\boldsymbol{Y}_t^2)$ denotes the element-wise logarithm of the squared observations and provides an observable proxy for the latent log conditional variance. The model jointly accommodates contemporaneous cross-sectional dependence, temporal persistence, and systematic variation associated with observed covariates, thereby allowing conditional variance dynamics to operate both across units and over time.

The matrix $\mathbf{W}=[w_{ij}]$ governs contemporaneous cross-sectional dependence, where $w_{ij}$ measures the contribution of unit $j$ to the conditional log-variance of unit $i$. In the financial application, a positive $w_{ij}$ indicates that higher contemporaneous log-squared returns for firm $j$ are associated with a higher conditional log-variance for firm $i$, conditional on the remaining components of the model. The matrix $\mathbf{W}$ is assumed to remain constant over the observation period. Accordingly, the estimated network should be interpreted as a time-invariant or average conditional dependence structure over the estimation window rather than as a dynamically evolving network. Allowing the interaction matrix to vary with time would require an additional dynamic specification for $\mathbf{W}_t$ and is left for future work.

The temporal autoregressive matrices $\mathbf{\Phi}_p$ are assumed to be diagonal, with $\mathbf{\Phi}_{p}=\operatorname{diag}\left(\phi_p(\boldsymbol{s}_1),\ldots,\phi_p(\boldsymbol{s}_n)\right)$ for $p=1,\ldots,P$, where $\phi_p(\boldsymbol{s}_i)=(\mathbf{\Phi}_p)_{ii}$ denotes the temporal autoregressive coefficient for unit $i$ at lag $p$. This specification allows temporal persistence to vary across units, while contemporaneous cross-sectional dependence is captured separately through $\mathbf{W}$. The matrix $\mathbf{X}_t$ contains exogenous covariates, with corresponding coefficient vector $\boldsymbol{\beta}$, which may include market-wide volatility measures, macroeconomic variables, or sector-specific indicators that account for systematic variation in the conditional variance. Finally, $\boldsymbol{\mu}=(\mu_1,\ldots,\mu_n)^\top$ denotes the vector of unit-specific intercepts, representing baseline log conditional variance levels after accounting for the contemporaneous, temporal, and covariate components of the model.

An important feature of the proposed model is that the contemporaneous dependence term involves the current observations through $\log(\boldsymbol{Y}_t^2)$, rendering the model implicit. Taking the element-wise logarithm of the squared observations in \eqref{eq: spmodel} and substituting the conditional variance specification in \eqref{eq:log volatility} yields
\begin{equation}
\log(\boldsymbol{Y}^2_t)=\mathbf{W}\log(\boldsymbol{Y}_t^2)+\sum_{p=1}^{P}\mathbf{\Phi}_p\log(\boldsymbol{Y}_{t-p}^2)+\mathbf{X}_t\boldsymbol{\beta}+\boldsymbol{\mu}+\log(\boldsymbol{\varepsilon}_t^2). \label{eq:logvolatility2}
\end{equation}
Provided that $(\mathbf{I}-\mathbf{W})$ is nonsingular, this expression admits the reduced-form representation
\begin{equation}
\log(\boldsymbol{Y}^2_t)=(\mathbf{I}-\mathbf{W})^{-1}\left[\sum_{p=1}^{P}\mathbf{\Phi}_p\log(\boldsymbol{Y}_{t-p}^2)+\mathbf{X}_t\boldsymbol{\beta}+\boldsymbol{\mu}+\log(\boldsymbol{\varepsilon}_t^2)\right]. \label{eq: reducedform}
\end{equation}
The multiplier $(\mathbf{I}-\mathbf{W})^{-1}$ captures the propagation of contemporaneous dependence throughout the system through both direct and indirect network interactions. Thus, a change associated with one unit may be transmitted beyond its immediate connections through successive interactions encoded by $\mathbf{W}$. The resulting conditional variance dynamics therefore reflect the combined effects of temporal persistence, observed covariates, and contemporaneous cross-sectional dependence.

Since the logarithm of the squared innovations has a non-zero expectation, the innovation term in \eqref{eq: reducedform} induces a constant shift in the conditional mean of the observable log-squared process. Under the Gaussian innovation assumption, $\varepsilon_{it}\sim N(0,1)$, it follows that $\varepsilon_{it}^2\sim\chi_1^2$ and
\begin{equation}
\mathbb{E}\{\log(\varepsilon_{it}^2)\}=-\gamma-\log(2),
\end{equation}
where $\gamma\approx0.5772$ denotes Euler's constant (see, e.g., \citealp{johnson1994continuous}). Hence, taking conditional expectations in \eqref{eq: reducedform}, with respect to the information set $\mathcal{F}_{t-1}$ available prior to observing $\boldsymbol{Y}_t$, gives
\begin{equation}
\mathbb{E}\!\left[\log(\boldsymbol{Y}_t^2)\mid\mathcal{F}_{t-1},\boldsymbol{X}_t\right]=(\mathbf{I}-\mathbf{W})^{-1}\left[\sum_{p=1}^{P}\mathbf{\Phi}_p\log(\boldsymbol{Y}_{t-p}^2)+\mathbf{X}_t\boldsymbol{\beta}+\boldsymbol{\mu}-\{\gamma+\log(2)\}\boldsymbol{1}_n\right]. \label{eq:conditional_mean_logY}
\end{equation}
Thus, forecasts of the observable log-squared process differ from those of the systematic log conditional variance component by the expected value of the log-squared innovation. Future innovations do not enter the conditional mean directly but contribute through the constant correction $-\{\gamma+\log(2)\}\boldsymbol{1}_n$. Consequently, the conditional mean depends on the temporal autoregressive component, the observed covariates, and the contemporaneous dependence structure represented by the multiplier $(\mathbf{I}-\mathbf{W})^{-1}$.

The reduced-form representation in \eqref{eq: reducedform} also highlights an important statistical challenge associated with estimating the spatial weights matrix. For a system comprising $n$ units, $\mathbf{W}$ contains $n(n-1)$ potential directed dependence parameters after excluding the diagonal elements. The number of unknown interactions therefore increases quadratically with the dimension of the system. For example, a network of twenty firms contains $20\times19=380$ possible directed links. Estimating all of these parameters without additional structural assumptions may lead to overparameterisation, unstable estimation, and poor out-of-sample performance.

Motivated by the expectation that only a subset of the potential interactions is relevant, we assume that the underlying spatial weights matrix is sparse, such that many elements of $\mathbf{W}$ are zero. In the financial application, this assumption implies that conditional-volatility dependence is concentrated among a subset of firms rather than being uniformly distributed across the market. Sparsity also facilitates interpretation by identifying the dominant dependence relationships while reducing the effective dimensionality of the model. These considerations motivate the use of penalised estimation. By imposing regularisation on the elements of $\mathbf{W}$, the proposed framework jointly estimates the model parameters and selects the non-zero elements of the dependence structure, yielding a parsimonious and interpretable network representation.

To ensure model identifiability and stability, the spatial weights matrix is constrained to satisfy
\begin{equation}
w_{ij}\geq0,\qquad w_{ii}=0,\qquad \sum_{j=1}^{n}w_{ij}<1,\qquad i=1,\ldots,n.
\end{equation}
The first constraint restricts the model to non-negative contemporaneous dependence, the second excludes self-interactions, and the third implies $\|\mathbf{W}\|_{\infty}<1$ and hence $\rho(\mathbf{W})<1$, where $\rho(\mathbf{W})$ denotes the spectral radius of $\mathbf{W}$. Consequently, $(\mathbf{I}-\mathbf{W})$ is nonsingular, ensuring the existence of the reduced-form representation in \eqref{eq: reducedform}. Related stability conditions are commonly imposed in spatial autoregressive models to obtain well-defined dependence structures \citep{lesage2009introduction}.

In conventional spatial and spatiotemporal models, the spatial weights matrix is typically specified a priori using external information such as geographical proximity, sectoral classifications, trade relationships, or correlation-based measures \citep{cressie2011statistics}. Such specifications may be difficult to justify in financial systems, where dependence may arise from complex and partially unobserved interactions among market participants \citep{forbes2002no,diebold2012better}. To address this limitation, the proposed framework treats $\mathbf{W}$ as an unknown parameter matrix and estimates its elements directly from the data. This data-driven approach allows the cross-sectional conditional-volatility dependence structure to be learned jointly with temporal persistence, covariate effects, and unit-specific baseline components.

\section{Parameter Estimation}\label{sec:estimation}
Estimating the spatial weights matrix $\mathbf{W}$ without structural assumptions poses presents a challenging high-dimensional inference problem. In particular, the number of potential spatial connections, $n(n-1)$, can be substantially larger than the available sample size $nT$, rendering conventional maximum likelihood estimation unstable or infeasible. To address this, we incorporate an $\ell_1$-penalisation framework within the maximum likelihood estimation procedure. The LASSO penalty imposes sparsity on the entries of $\mathbf{W}$, shrinking weak or statistically insignificant connections towards zero while retaining the most relevant dependence relationships \citep{tibshirani1996regression}. This regularisation approach is consistent with recent developments in network inference \citet{meinshausen2006high} and spatial econometrics \cite{otto2023estimation},where dependence structures are learned directly from the data rather than imposed a priori. Consequently, the proposed framework allows both temporal and cross-sectional volatility dependencies to emerge naturally from the observed data.

The unknown parameters are collected in $\Theta =
\left\{
\mathbf{W},\mathbf{\Phi}_1,\ldots,\mathbf{\Phi}_P,
\boldsymbol{\beta},\boldsymbol{\mu},\mathbf{\Sigma}
\right\}$, 
where $\mathbf{W}$ is the contemporaneous spillover matrix, 
$\mathbf{\Phi}_1,\ldots,\mathbf{\Phi}_P$ are the temporal autoregressive coefficient matrices, 
$\boldsymbol{\beta}$ denotes the regression coefficients, $\boldsymbol{\mu}$ contains the spatial fixed effects, 
and $\mathbf{\Sigma}$ is the covariance matrix of the transformed innovations. For
$$\boldsymbol{\Upsilon}_t(\Theta)=(\mathbf{I}_n-\mathbf{W})\log(\boldsymbol{Y}_t^2)
-\sum_{p=1}^{P}\mathbf{\Phi}_p\log(\boldsymbol{Y}_{t-p}^2)-\mathbf{X}_t\boldsymbol{\beta}-\boldsymbol{\mu},$$
a Gaussian quasi log-likelihood for the transformed process is given by
{\small
\begin{equation}
\ell(\Theta) = -\frac{n(T-P)}{2}\log(2\pi) -\frac{T-P}{2}\log|\mathbf{\Sigma}| + (T-P)\log|\mathbf{I}_n-\mathbf{W}| - \frac{1}{2} \sum_{t=P+1}^{T}\boldsymbol{\Upsilon}_t(\Theta)^\top \mathbf{\Sigma}^{-1} \boldsymbol{\Upsilon}_t(\Theta).\label{eq:loglikelihood}
\end{equation}
}
The determinant term arises from the Jacobian of the spatial autoregressive transformation 
from $\log(\boldsymbol{Y}_t^2)$ to 
$(\mathbf{I}_n-\mathbf{W})\log(\boldsymbol{Y}_t^2)$ and is standard in likelihood-based estimation of spatial autoregressive models \citep{lesage2009introduction}. The quadratic term measures the discrepancy between the transformed observed log-squared process and its fitted conditional mean under the Gaussian quasi-likelihood approximation.

Since the number of possible directed spillover coefficients in $\mathbf{W}$ increases as $n(n-1)$, direct unregularised estimation is unstable in moderate or high dimensions. We therefore impose sparsity through an $\ell_1$-penalised quasi-likelihood. The penalised estimator is defined as
\begin{align}
\widehat{\Theta}=\arg\min_{\Theta}\;&\frac{T-P}{2}\log|\mathbf{\Sigma}|-(T-P)\log|\mathbf{I}_n-\mathbf{W}|+
\frac{1}{2}\sum_{t=P+1}^{T}\boldsymbol{\Upsilon}_t(\Theta)^\top\mathbf{\Sigma}^{-1}\boldsymbol{\Upsilon}_t(\Theta) \nonumber\\
&\quad+\lambda_1\sum_{i\neq j}|w_{ij}|+\lambda_2\sum_{p=1}^{P}\|\mathbf{\Phi}_p\|_1+ \lambda_3\sum_{k=1}^{K}|\beta_k|,\label{eq:penalised_objective}
\end{align}
subject to
$$w_{ij}\geq 0,\qquad w_{ii}=0,\qquad \sum_{j=1}^{n}w_{ij}\leq 1,\qquad i=1,\ldots,n.$$
Here, $\lambda_1,\lambda_2,\lambda_3\geq0$ control the degree of shrinkage applied to the spatial spillover matrix, the temporal autoregressive coefficients, and the covariate effects, respectively. The row-sum constraint bounds the total contemporaneous influence received by each unit and ensures that $\mathbf{I}_n-\mathbf{W}$ remains nonsingular under the stated non-negativity restriction. The $\ell_1$ penalty on $\mathbf{W}$ produces a sparse directed volatility network by shrinking weak spillover coefficients to zero, while the penalties on $\mathbf{\Phi}_p$ and $\boldsymbol{\beta}$ regularise the temporal and covariate effects. The optimisation is carried out by constrained numerical minimisation of \eqref{eq:penalised_objective}. 

This present study focuses on finite-sample estimation and recovery of the dependence structure rather than establishing a high-dimensional asymptotic theory for the penalised estimator. In particular, formal results on parameter consistency, support-selection consistency, convergence rates, or asymptotic distributions are not derived here. The joint presence of an unknown contemporaneous interaction matrix, temporal conditional heteroskedasticity, structural constraints, and regularisation makes these questions non-trivial. The finite-sample behaviour of the estimator is therefore investigated systematically through the Monte Carlo experiments in Section~\ref{sec:simulationstudy}, where both coefficient-estimation accuracy and recovery of the sparsity pattern are evaluated. 

\subsection{Selection of the Regularisation Parameters}\label{sec:tuning_parameters}

The penalised objective in \eqref{eq:penalised_objective} contains three tuning parameters, $\lambda_1$, $\lambda_2$, and $\lambda_3$, controlling the regularisation applied to $\mathbf{W}$, $\mathbf{\Phi}_p$, and $\boldsymbol{\beta}$, respectively. Each parameter is selected from the grid $\Lambda_j=\{0\}\cup\{10^{-2},10^{-1.5},10^{-1},10^{-0.5},10^{0},10^{0.5},10^{1},10^{1.5},10^{2}\}$, for $j=1,2,3$, giving $10^3=1000$ candidate combinations over $\Lambda=\Lambda_1\times\Lambda_2\times\Lambda_3$. The inclusion of zero allows the corresponding component to remain unpenalised.

To preserve the temporal ordering of the data, tuning is performed using an expanding-window cross-validation procedure. The estimation sample is divided chronologically into five consecutive blocks, with the first block forming the initial training sample and each subsequent block serving as a validation period after fitting the model to all preceding observations. This produces four validation folds and prevents future observations from entering the corresponding training samples. For a candidate penalty vector $\boldsymbol{\lambda}=(\lambda_1,\lambda_2,\lambda_3)^\top$, the overall validation criterion is
\begin{equation}
CV(\boldsymbol{\lambda})=\frac{1}{R_{\mathrm{CV}}}\sum_{r=1}^{R_{\mathrm{CV}}}L_r(\boldsymbol{\lambda}), \qquad R_{\mathrm{CV}}=4,
\label{eq:cv_penalty_selection}
\end{equation}
and the selected penalty combination is
\begin{equation}
\widehat{\boldsymbol{\lambda}}=\underset{\boldsymbol{\lambda}\in\Lambda}{\arg\min}\;CV(\boldsymbol{\lambda}).
\label{eq:selected_lambda}
\end{equation}

The validation loss $L_r(\boldsymbol{\lambda})$ depends on the analysis. In the Monte Carlo study, where the latent conditional variance is known, tuning is based on prediction of the true log conditional variance. In the empirical application, where the conditional variance is unobserved, tuning is based on the observable log-squared return process. The corresponding validation criteria are specified in Sections~\ref{sec:simulationstudy} and~\ref{sec:stock_application}, respectively. After selecting $\widehat{\boldsymbol{\lambda}}$, the model is re-estimated on the complete estimation sample using the selected penalty combination.

\section{A Monte Carlo Study}\label{sec:simulationstudy}
A Monte Carlo simulation study was conducted to evaluate the finite-sample performance of the proposed penalised estimator in recovering the regression coefficients, temporal dependence parameters, spatial spillover structure, and latent volatility components. Particular emphasis was placed on assessing the ability of the method to accurately recover sparse volatility networks under varying spatial and temporal dimensions. Data were generated from the spatiotemporal ARCH model introduced in Section \ref{sec:ARCH_Model}. Specifically, observations were generated according to \eqref{eq: spmodel} where innovations were independently generated from a standard normal distribution, $\boldsymbol{\varepsilon}_t \sim N(\mathbf{0},\mathbf{I}_n)$. The latent log-volatility process was generated recursively from \eqref{eq:log volatility}. Using the observation \eqref{eq: spmodel}, this may equivalently be simulated through the reduced-form representation of \eqref{eq: reducedform}. This representation makes clear that contemporaneous spatial spillovers are transmitted through the spatial multiplier $(\mathbf{I}-\mathbf{W})^{-1}$. 

The regression coefficient vector, $\boldsymbol{\beta}$, was set to $(3, 0, 2)^\top$, where the second coefficient was set to zero to assess the ability of the penalised estimator to identify and remove irrelevant covariates through shrinkage. The corresponding covariates were generated independently from a standard normal distribution, such that,
\begin{equation*}
x_{ijt} \sim \mathcal{N}(0,1), \qquad i=1,\ldots,n, \, t=1,\ldots,T+100, \, j=1,\ldots,k,
\end{equation*}
where $i$ indexes the spatial unit, $t$ the time point, and $j$ the covariate. Consequently, the covariates were independent across both space and time with zero mean and unit variance. An additional $100$ observations were generated as a burn-in period to reduce the influence of the initial conditions before retaining the final T observations for analysis. The temporal dependence matrix was specified as a diagonal matrix, both zero and non-zero elements, allowing assessment of both active and inactive coefficients. The temporal persistence parameters were set such that the first $25\%$ of the spatial units exhibit no temporal dependence, while the remaining $75\%$ share a uniform autoregressive coefficient of $\phi= 0.6$ with a time lag of $P=1$. Spatiotemporal diffusion effects, i.e., non-zero off-diagonal elements in $(\mathbf{\Phi}_1,\ldots,\mathbf{\Phi}_P)$, were not considered. This design introduces heterogeneity in temporal volatility persistence across spatial units and provides a useful setting for assessing whether the proposed estimator can distinguish persistent from non-persistent components. The spatial spillover matrix $\mathbf{W}$ was constructed using a Queen-contiguity neighbourhood structure on a regular lattice. The resulting adjacency matrix was row-standardised and then multiplied by a spatial dependence parameter $\rho=0.4$. This produces a sparse spatial spillover structure in which each unit is directly connected only to its immediate neighbours. The diagonal elements were set to zero to exclude self-dependence through the spatial component. Spatial fixed effects were generated from  $\boldsymbol{\mu_i} \sim \mathcal{N}(0,1).$ To reduce the influence of initial conditions, a burn-in period of 100 observations was discarded before estimation.

The simulation study considered combinations of spatial dimensions, $ n \in\{4, 9, 16,25\} $ and temporal sample sizes $T \in \{50, 100, 200\}$, leading to observations-to-parameter ratios ranging from $1.91$ to $34.78$. For each $(n, T)$ configuration, $R=100$ independent Monte Carlo replications were generated, allowing evaluation of how estimator performance changes as both the cross-sectional and temporal dimensions increase. For each replication, the model parameters were estimated using the proposed LASSO-penalised quasi-maximum likelihood procedure. The regularisation parameters,  $\lambda_1, \lambda_2, \lambda_3$ associated with the regression coefficients, temporal persistence parameters, and spatial spillover matrix were selected using five-fold cross-validation. The spatial weighting matrix represents contemporaneous cross-sectional interactions and is assumed constant throughout the observation period. Consequently, cross-validation was performed by partitioning the data only along the temporal dimension while preserving the full spatial configuration within each fold. This ensures that the spatial dependence structure is estimated using complete cross-sectional information while avoiding temporal information leakage. For each candidate tuning-parameter combination, the model was estimated on the training blocks and evaluated on the validation block. The optimal penalty combination was selected by minimising the average out-of-sample root mean squared prediction error across the cross-validation folds, 
\begin{equation*}
\mathrm{CV}(\lambda_1,\lambda_2,\lambda_3)=\left[\frac{1}{n(T-1)}\sum_{t=2}^{T}\sum_{i=1}^{n}\left(
\widehat{\log h}_t(s_i)-\log h_t(s_i)\right)^2\right]^{1/2}.\label{eq:cv_rmse}
\end{equation*}
Thus, the tuning parameters were chosen exclusively using out-of-sample prediction performance. After selecting the optimal tuning parameters, the model was re-estimated using the complete dataset. The finite-sample performance of the resulting estimator was then evaluated over the Monte Carlo replications using the bias, mean absolute error (MAE), and root mean squared error (RMSE). Since these measures were computed from the parameter estimates obtained after refitting the model to the full dataset, they assess the estimation accuracy of the final fitted model rather than out-of-sample predictive performance. For a generic parameter $\theta$, these measures were computed as
\begin{equation*}
\text{Bias} = \frac{1}{R} \sum_{r=1}^{R} \left( \hat{\theta}^{(r)} - \theta \right), \\
\text{MAE} = \frac{1}{R} \sum_{r=1}^{R} \left| \hat{\theta}^{(r)} - \theta \right|, \\
\text{RMSE} = \sqrt{\frac{1}{R} \sum_{r=1}^{R} \left( \hat{\theta}^{(r)} - \theta \right)^2}.
\end{equation*}
These measures were evaluated separately for the regression coefficients $\boldsymbol{\beta}$, temporal persistence parameters $\boldsymbol{\Phi}$, spatial fixed effects $\boldsymbol{\mu}$, and spatial weights $\mathbf{W}$. For the spatial weights matrix, performance was additionally summarised separately for zero and non-zero elements of $\mathbf{W}$. This distinction is important because the proposed method is designed not only to estimate the magnitude of non-zero spillovers but also to recover the sparse network structure.  Computational time was also recorded for each simulation configuration to assess the scalability of the estimation procedure.

\noindent Table \ref{tab: results} reports the average bias, MAE, and RMSE across all simulation settings. Overall, the proposed estimator performs well across the considered spatial and temporal dimensions. Estimation accuracy improves systematically as the temporal sample size increases, indicating that longer time series provide more information for recovering both temporal persistence and contemporaneous spatial spillovers.
\begin{sidewaystable}[!p]
\caption{Average Bias, Mean Absolute Error and Root Mean Square Error (RMSE) of Estimated Parameters}
\label{tab: results}
\begin{tabular}{cccccccccccccc}
\hline
\multirow{3}{*}{Parameter} & \multirow{3}{*}{Measure} & \multicolumn{12}{c}{Spatiotemporal ARCH Model performance metrics } \\\cline{3-14}
                           &                        & \multicolumn{3}{c}{$n = 4$}  & \multicolumn{3}{c}{$n = 9 $} & \multicolumn{3}{c}{$n = 16 $}  & \multicolumn{3}{c}{$n = 25 $}  \\\cline{3-14} &                          
& $T = 50$     &  $100$  & $200 $  & $50$     &  $100$  & $200$ & $50$     &  $100$  & $200 $   & $50$     &  $100$  & $200 $  \\\hline
\multirow{2}{*}{Beta ($\boldsymbol{\beta}$)}      
& Bias    & 0.0035 & 0.0053  & 0.0039& 0.003 &0.0034 & 0.0094 & 0.0128  & 0.0089 &0.0051  & 0.0184 &0.009 &0.0058 \\
& MAE    &  0.133 & 0.0888  &  0.0655 & 0.1015   & 0.0648 & 0.0437 &  0.076 & 0.0491   & 0.0308  & 0.0553  & 0.0406     & 0.0264  \\
& RMSE    & 0.1667 & 0.1115 & 0.0825 & 0.1228 & 0.0822 & 0.0549 &  0.095 & 0.0594 & 0.0392  &0.0697  &0.0502 &0.0338 \\
\hline
\multirow{2}{*}{Phi ($\mathbf{\Phi}$)}        
& Bias    & -0.0028 &0.002  &0.0002 & -0.0067 & -0.0041& 0.0002 &-0.0085  & -0.0048 & 0.0003 & -0.0125 &-0.0044 &0.0002  \\
& MAE    &   0.043  & 0.0301  & 0.0186 &  0.0441   & 0.0326 &0.0221  & 0.0501 & 0.0356   &0.0253   & 0.0531 & 0.036    &0.0251   \\
& RMSE   &  0.0582 & 0.0399 & 0.0245& 0.0593 & 0.0429 & 0.0298  &0.0664  &0.0461 & 0.0315  & 0.0702 &0.0464 &0.0316 \\
\hline
\multirow{2}{*}{$\mathbf{W}$}           
& Bias         & -0.0021 &-0.0011  & -0.0009& 0.0034 &0.0018 &0.0013  &0.0055  & 0.0038 & 0.0026  & 0.0071  & 0.0046 & 0.0033 \\
& MAE    & 0.0686  & 0.0422  &   0.0299     & 0.0423   & 0.0291 & 0.0205 & 0.03 &  0.0216   & 0.0153  & 0.0239  & 0.0171 &0.0123   \\
& RMSE    & 0.0859 & 0.0536 & 0.0379& 0.0589 & 0.0413 &0.029  &0.0468  & 0.0327 &0.0228  &0.0406  &0.0279 &0.0194 \\
\hline
\multirow{2}{*}{$\mathbf{W} = 0$}           
& Bias  &  &  & &  0.0205 & 0.0138&0.0093  &0.0153   &0.0107 &0.0077  &0.0136  & 0.0094 &0.0068 \\
& MAE   & -- & -- & -- & 0.0205   & 0.0138 & 0.0093 &0.0153 &0.0107    &0.0077   &0.0136  & 0.0094    & 0.0068  \\
& RMSE  &  &  & & 0.0416  &0.0285 &0.0188  &0.0351  &0.0229 &0.0154  & 0.0322 & 0.0209 & 0.0139 \\
\hline
\multirow{2}{*}{$\mathbf{W} \neq 0$ }        
& Bias     & -0.0021 & -0.0011 &-0.0009 & -0.0104 &-0.0079 & -0.0051 &-0.0127  &  -0.0089 &-0.0068  &-0.0135  &-0.0104 &-0.0076 \\
& MAE    & 0.0686 &0.0422 & 0.0299 & 0.0598 &0.0413 & 0.0295 & 0.0571  &0.0419 &0.0296   & 0.0566  &0.0415 &0.0297 \\
& RMSE     & 0.0859 & 0.0536 & 0.0379&0.0728  &0.0515 & 0.037 &0.0686  &0.0508 & 0.0366 & 0.0674  &0.0502 &0.0366 \\
\hline
\multirow{2}{*}{$\boldsymbol{\mu}$}    
& Bias  &  0.0128 & 0.0039 & 0.0033&0.0256  &0.0015  &0.0115  &0.1014  & 0.0682 &  0.0551 &0.2619  & 0.1788 &0.1264 \\
& MAE    & 0.292  & 0.1781 & 0.1276 &0.3112  & 0.2134  & 0.1423 & 0.3531 & 0.2326  &0.1583  &0.4575  &0.2796 &0.1921 \\
& RMSE  & 0.3712 & 0.2298 & 0.1591 & 0.3913 &0.2657 & 0.1823 & 0.447 & 0.2902 & 0.1954 &0.5718  &0.3464 &0.2372 \\
\hline
Time $t$        
&  time(secs)   & 354.00 & 341.70 & 666.00 & 4862.2 &9003.7 & 17842  &23038  & 36083 &  61914 &98244.1  &128692 &190301 \\\hline
                      
\end{tabular}
\end{sidewaystable}
The regression coefficients are estimated with very small bias across all configurations. Both MAE and RMSE decline as $T$ increases, demonstrating that the estimator accurately recovers the covariate effects even in the presence of spatial and temporal volatility dependence. The inactive coefficient is also shrunk close to zero, illustrating the variable-selection capability of the LASSO penalty. The temporal autoregressive parameter $\mathbf{\Phi}$ also exhibits low bias across all scenarios, with MAE and RMSE decreasing as $T$ increases, reflecting the model’s improved ability to capture temporal dependence with longer time series. 
The temporal persistence parameters are also recovered accurately. Bias remains close to zero in all scenarios, while MAE and RMSE decrease as the temporal dimension increases indicating that the estimator performs well with increasing time resolutions. 
\begin{figure}[!t]
\centering
\includegraphics[width=\linewidth]{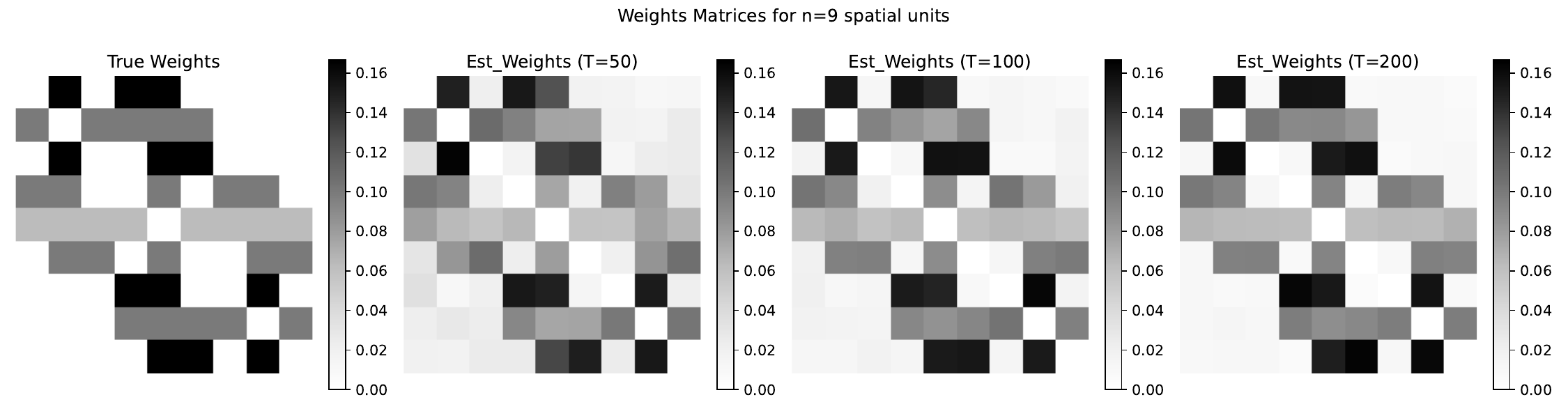}
\caption{Estimated spatial weights for 9 spatial units}
\label{fig: W9}
\end{figure}

The estimation of the spatial weights matrix is more challenging because the number of potential spillover parameters grows quadratically with the number of spatial units. Nevertheless, the proposed method successfully recovers the underlying spatial dependence structure. The overall bias of $\mathbf{W}$ remains small, and both MAE and RMSE decrease as $T$ increases. This confirms that longer temporal samples improve the recovery of contemporaneous spillover effects. The estimated spatial weights for each time point for nine spatial units are shown in the Figure \ref{fig: W9}. 

The separate evaluation of zero and non-zero spatial weights provides further insight into network recovery. Estimation errors for zero weights are small, suggesting that the penalisation procedure is effective in shrinking irrelevant spillover links. Non-zero weights are more difficult to estimate, as expected, but their errors decrease with increasing $T$. This pattern indicates that the method is able to recover both the sparsity pattern and the magnitude of the dominant spatial spillovers. The spatial fixed effects $\boldsymbol{\mu}$ are more difficult to estimate than the regression and dependence parameters. Their MAE and RMSE are comparatively larger, particularly as the spatial dimension increases. This reflects the difficulty of estimating unit-specific baseline volatility levels in a high-dimensional conditional heteroskedastic model. However, estimation accuracy improves with larger $T$, indicating that longer time series mitigate this difficulty.

The computational results show that estimation time increases substantially with both $n$ and $T$. This increase is expected because the number of spatial weights grows as $n(n-1)$, and each cross-validation step requires repeated constrained optimisation. The results, therefore, highlight the computational cost of jointly estimating temporal dependence, covariate effects, spatial fixed effects, and a full spatial spillover matrix. Nevertheless, the simulation study demonstrates that the proposed method is feasible for moderate-dimensional systems and provides accurate recovery of the underlying spatiotemporal volatility structure. 
The Monte Carlo results provide evidence that the proposed LASSO-penalised Spatiotemporal ARCH estimator performs well in finite samples. Across all simulation configurations, the estimator produces low bias for the regression, temporal, and spatial dependence parameters. The systematic decline in MAE and RMSE as $T$ increases is consistent with improved estimation accuracy under larger temporal samples.

Moreover, the results also support the central methodological motivation of the paper: sparse regularisation is useful for learning unknown volatility spillover networks. By shrinking weak or irrelevant entries of $\mathbf{W}$ towards zero, the proposed estimator produces parsimonious and interpretable dependence structures while retaining the dominant spatial interactions. The simulation study demonstrates that the proposed estimator can recover complex spatiotemporal volatility dynamics under moderate sample sizes. The method is particularly effective when sufficient temporal observations are available, although computational cost remains an important consideration for larger systems. These findings motivate the empirical application in the next section, where the proposed framework is applied to volatility spillovers among UK-listed firms.

\section{Volatility Spillovers in UK Equity Markets}\label{sec:stock_application}
We apply the proposed method to daily returns from the London Stock Exchange (LSE). Financial returns exhibit both temporal volatility persistence and cross-sectional dependence arising from common market conditions, sectoral links and other economic relationships. Because these connections are rarely observed directly, predefined networks based on sector membership or other external information may not adequately represent volatility dependence. The application therefore examines whether this structure can be learned directly from financial returns and whether doing so improves volatility prediction.

\subsection{Data}
We analyse twenty UK-listed firms from four broad groups: Banking and Financial Services, Energy and Resources, Consumer Goods, and Telecommunications and Pharmaceuticals. The firms were selected primarily for data completeness over the study period, limiting the need for imputation and enabling construction of a balanced panel. Subject to this criterion, firms from different sectors were retained to examine both within-sector and cross-sector dependence. Table \ref{Stocks_info} lists the firms and their classifications.
\begin{table}[!t]
\centering
\small
\begin{spacing}{1}
\caption{FTSE 100 Selected Stocks by Sector with Description}
\label{Stocks_info}
\begin{tabular}{p{3cm}p{2cm}p{3.5cm}p{7cm}}
\hline
\textbf{Stock Name} & \textbf{Ticker} & \textbf{Category} & \textbf{Description} \\
\hline
Barclays & BARC.L & Banking/Financials & Major UK bank, retail and corporate banking. \\
HSBC Holdings & HSBA.L & Banking/Financials & Global banking and financial services provider. \\
Lloyds Group & LLOY.L & Banking/Financials & Leading UK retail and commercial bank. \\
Prudential & PRU.L & Banking/Financials & Insurance and financial services company. \\
London Stock EG & LSEG.L & Banking/Financials & Operates financial markets and data services \\
\hline
Shell & SHEL.L & Energy/Resources & Multinational oil and gas company. \\
National Grid & NG.L & Energy/Resources & Transmits electricity and gas in the UK. \\
Rio Tinto & RIO.L & Energy/Resources & Global mining and metals company. \\
Anglo American & AAL.L & Energy/Resources & Diversified mining company. \\
Glencore & GLEN.L & Energy/Resources & Mining and commodity trading group. \\
\hline
Unilever & ULVR.L & Consumer Goods & Global consumer goods and food products company \\
Diageo & DGE.L & Consumer Goods & Leading producer of alcoholic beverages. \\
British A Tobacco & BATS.L & Consumer Goods & Multinational tobacco and nicotine products firm. \\
Tesco & TSCO.L & Consumer Goods & Major UK supermarket and retail chain. \\
Coca-Cola AG & CCH.L & Consumer Goods & Bottles and distributes Coca-Cola products. \\
\hline
Vodafone Group & VOD.L & Telecom \& Pharma & Multinational telecoms group. \\
AstraZeneca & AZN.L & Telecom \& Pharma & Global biopharmaceutical company. \\
GlaxoSmithKline & GSK.L & Telecom \& Pharma & Pharmaceutical and healthcare company. \\
RELX & REL.L & Telecom \& Pharma & Information, analytics and publishing firm. \\
Informa & INF.L & Telecom \& Pharma & Organizes exhibitions and business intelligence. \\
\hline
\end{tabular}
\end{spacing}

\end{table}
Daily closing prices were obtained from Yahoo Finance for January 2024 to December 2025. For firm $i=1,\ldots,n$ and time $t=2,\ldots,T$, the daily log return is
\begin{equation}
    y_{i,t} =\log\left(\frac{P_{i,t}}{P_{i,t-1}}\right),\label{eq:log_returns_stock}
\end{equation}
where $P_{i,t}$ is the closing price. The returns are collected as $\mathbf{Y}_t=(y_{1,t},\ldots,y_{n,t})^\top\in\mathbb{R}^n$, with $n=20$. The data contained six missing returns: four for Glencore and two for Diageo. These resulted from three missing prices because each missing price affects both the current and subsequent return. As the missing returns represented less than $0.1\%$ of the sample, the affected dates were removed across all firms, producing a balanced panel with $T=500$ trading days. This sample provides sufficient temporal information for estimation, cross-validation and out-of-sample evaluation while keeping the joint estimation of the parameters and $\mathbf{W}$ computationally manageable.

\begin{figure}[!t]
    \centering
    \includegraphics[width=0.65\linewidth]{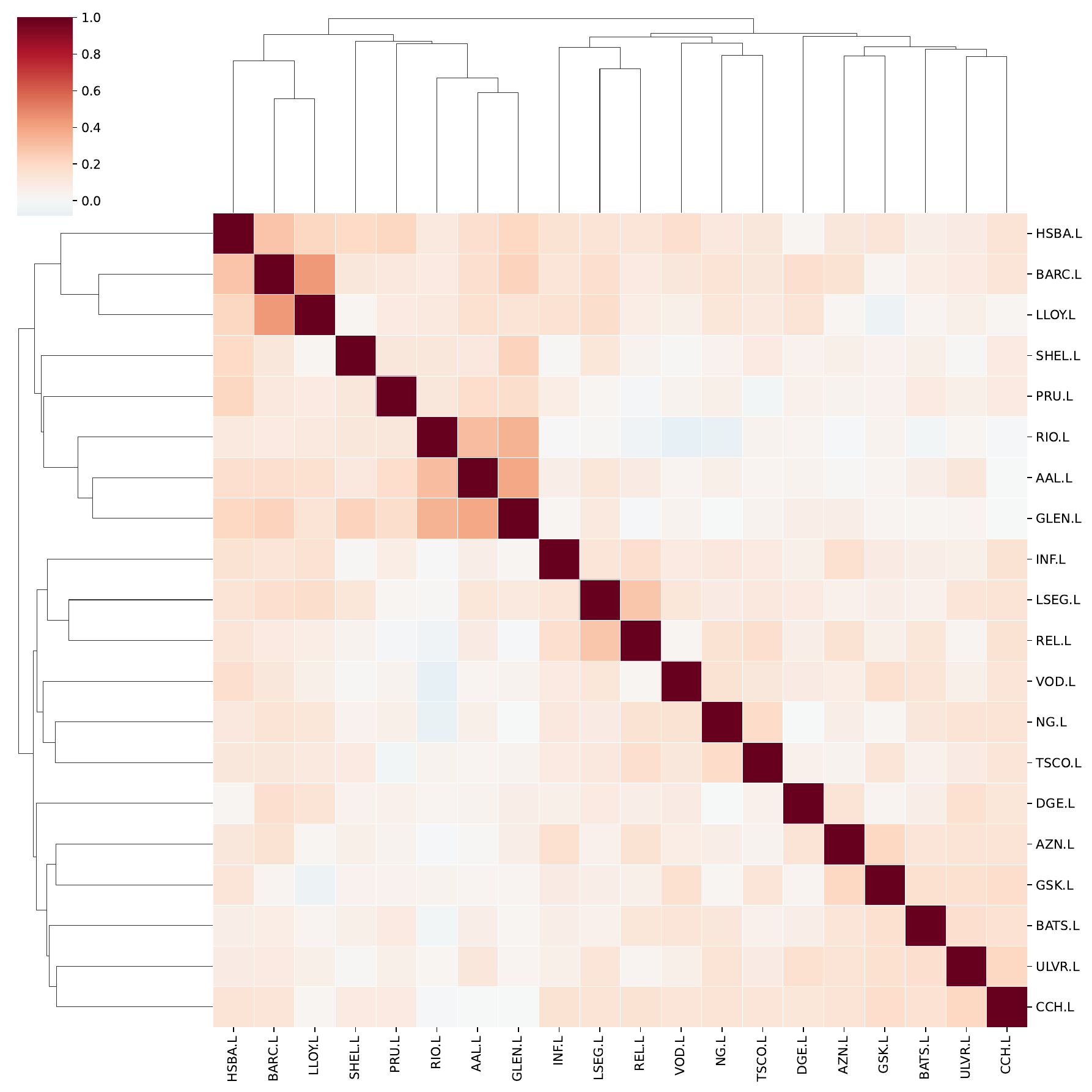}
    \caption{Clustered correlation matrix of log-squared returns for the selected FTSE-listed firms. Firms with similar volatility dynamics are grouped using hierarchical clustering.}
    \label{fig:cluster_corr_heatmap}
\end{figure}

\begin{table}[!htbp]
\centering
\caption{Summary statistics of daily log returns for the selected FTSE-listed firms over the sample period. Annual volatility is computed as $\hat{\sigma}\sqrt{252}$, where $\hat{\sigma}$ denotes the sample standard deviation of daily returns.}
\label{tab:summary_statistics}
\resizebox{\textwidth}{!}{
\begin{tabular}{lrrrrrrrr}
\hline
Company  & Mean & Std. Dev. & Min & Median & Max & Skewness & Kurtosis & Annual Vol. \\
\hline
Barclays & 0.0006 & 0.0236 & -0.1901 & 0.0009 & 0.1473 & -0.3311 & 10.3059 & 0.3751 \\
HSBC Holdings & 0.0002 & 0.0175 & -0.1001 & 0.0008 & 0.1018 & -0.3377 & 7.7874 & 0.2781 \\
Lloyds Group & 0.0002 & 0.0209 & -0.1297 & 0.0002 & 0.1176 & -0.2018 & 8.6193 & 0.3317 \\
Prudential & -0.0003 & 0.0239 & -0.1825 & 0.0001 & 0.1644 & -0.1731 & 10.0125 & 0.3798 \\
London Stock EG & 0.0003 & 0.0169 & -0.1552 & 0.0007 & 0.1427 & -0.1953 & 15.2451 & 0.2680 \\
Shell & 0.0000 & 0.0207 & -0.1935 & 0.0006 & 0.1855 & -0.6327 & 17.5577 & 0.3287 \\
National Grid & 0.0002 & 0.0142 & -0.1150 & 0.0010 & 0.0935 & -0.9317 & 11.9807 & 0.2254 \\
Rio Tinto & -0.0000 & 0.0189 & -0.1254 & 0.0004 & 0.1371 & -0.0871 & 7.5591 & 0.3004 \\
Anglo American & -0.0000 & 0.0268 & -0.2103 & 0.0002 & 0.1898 & -0.4415 & 10.8326 & 0.4260 \\
Glencore & 0.0001 & 0.0236 & -0.1887 & 0.0006 & 0.1291 & -0.6293 & 7.9887 & 0.3747 \\
Unilever & -0.0001 & 0.0128 & -0.0743 & 0.0000 & 0.0938 & 0.2505 & 11.5029 & 0.2028 \\
Diageo & -0.0004 & 0.0151 & -0.1298 & -0.0001 & 0.0939 & -0.3590 & 10.9719 & 0.2399 \\
British A Tobacco & 0.0003 & 0.0152 & -0.1091 & 0.0008 & 0.0732 & -0.6348 & 8.4698 & 0.2416 \\
Tesco & 0.0003 & 0.0145 & -0.2209 & 0.0005 & 0.0631 & -2.7002 & 41.1433 & 0.2298 \\
Coca-Cola AG & 0.0002 & 0.0182 & -0.1624 & 0.0003 & 0.1342 & -0.5656 & 16.0299 & 0.2888 \\
Vodafone Group & -0.0003 & 0.0178 & -0.1225 & 0.0001 & 0.1008 & -0.3983 & 8.6680 & 0.2830 \\
AstraZeneca & 0.0004 & 0.0158 & -0.0967 & 0.0006 & 0.0773 & -0.3191 & 7.2771 & 0.2516 \\
GlaxoSmithKline & -0.0001 & 0.0143 & -0.1060 & 0.0005 & 0.0733 & -0.6566 & 9.4882 & 0.2278 \\
RELX & 0.0005 & 0.0137 & -0.1143 & 0.0011 & 0.0915 & -0.4708 & 10.4785 & 0.2169 \\
Informa & 0.0000 & 0.0208 & -0.1098 & 0.0000 & 0.2003 & 0.2610 & 12.5722 & 0.3299 \\
\hline
\end{tabular}}
\end{table}
Table \ref{tab:summary_statistics} shows that mean daily returns are close to zero, while volatility varies considerably across firms. Anglo American, Prudential, Barclays and Glencore have the largest standard deviations. Most returns are negatively skewed, and kurtosis ranges from $7.28$ to $41.14$, indicating heavy tails. These features motivate a conditional heteroskedasticity model with heterogeneous firm-level volatility dynamics.
\begin{figure}[!t]
    \centering
    \includegraphics[width=\linewidth]{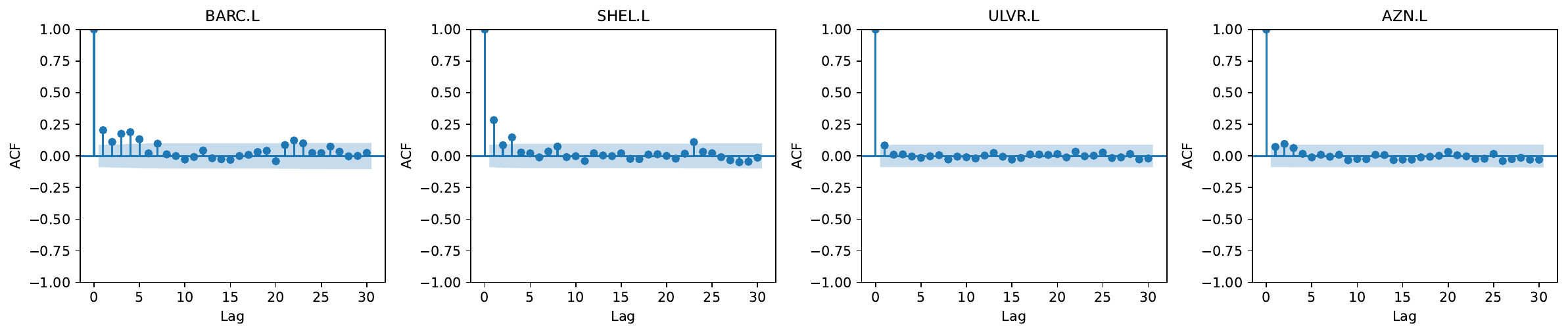}
    \caption{Autocorrelation functions of squared returns for selected stocks.}
    \label{fig:acf_squared_log_returns}
\end{figure}
Figure \ref{fig:acf_squared_log_returns} shows heterogeneous persistence in squared returns. Barclays and Shell exhibit positive autocorrelation at several lags, whereas Unilever and AstraZeneca show weaker serial dependence. This variation supports firm-specific temporal persistence parameters. Figure~\ref{fig:cluster_corr_heatmap} indicates cross-sectional association among the firms’ log-squared returns, including relationships within the financial and resource sectors. However, correlation is symmetric and cannot identify directional conditional dependence. This motivates estimating $\mathbf{W}$ directly from the data.

\subsection{Learning the Dependence Structure}

To investigate these dependence structures, the panel of daily log returns is analysed using the penalised Spatiotemporal ARCH model introduced in Section \ref{sec:ARCH_Model}. Let $\boldsymbol{Y}_t=\left(Y_t(\boldsymbol{s}_1),\ldots,Y_t(\boldsymbol{s}_n)\right)^\top$ denote the vector of stock returns across the $n$ firms at time $t$. Conditional on the variance process, the returns follow the specification in \eqref{eq: spmodel}, while the log-volatility process evolves according to \eqref{eq:log volatility}. For the empirical application, we consider a first-order temporal specification ($P=1$), such that $\mathbf{\Phi}_1=\operatorname{diag}\{\phi(\boldsymbol{s}_1),\ldots,\phi(\boldsymbol{s}_n)\}$. In the empirical application, no exogenous covariates are included. Accordingly, $\mathbf{X}_t$ is reduced to a constant term with a single coefficient $\beta$, which represents the common intercept across firms, while $\boldsymbol{\mu}$ captures firm-specific deviations in baseline log conditional variance. Thus, the empirical specification estimates one common intercept, firm-specific temporal persistence parameters, the unit-specific baseline effects, and the contemporaneous dependence matrix $\mathbf{W}$. Particular attention is given to the estimated dependence matrix $\mathbf{W}=(w_{ij})_{i,j=1}^{n}$, where $w_{ij}$ quantifies the contemporaneous conditional-volatility dependence of firm $i$ on firm $j$. Unlike conventional spatial or network specifications in which the dependence structure is defined a priori using geographical proximity, sectoral classifications, or other external information, $\mathbf{W}$ is estimated directly from the observed return series using the proposed LASSO-penalised quasi-maximum likelihood procedure. The resulting matrix therefore provides a data-driven representation of the cross-sectional conditional-volatility dependence structure among firms. Because the empirical model contains no covariates beyond an intercept, the intercept was left unpenalised, and no regularisation parameter was selected for $\beta$. The remaining regularisation parameters, $(\lambda_{\phi},\lambda_W)$, were selected by cross-validation using prediction RMSE. Rather than choosing the combination with the absolute minimum cross-validation RMSE, we applied the one-standard-error rule. Candidate combinations were evaluated over predefined search grids, and all combinations with a mean cross-validation RMSE within one standard error of the minimum were considered admissible. Among these combinations, we selected the one with the largest $\lambda_W$ to favour a sparser dependence matrix while retaining predictive performance comparable to that of the minimum-error model. The model was subsequently re-estimated using the selected $(\lambda_{\phi},\lambda_W)$, and the resulting estimates are reported in the following subsections.

\subsection{Estimated Volatility Dependence Network}
\begin{figure}[!t]
    \centering
    \includegraphics[width=\linewidth]{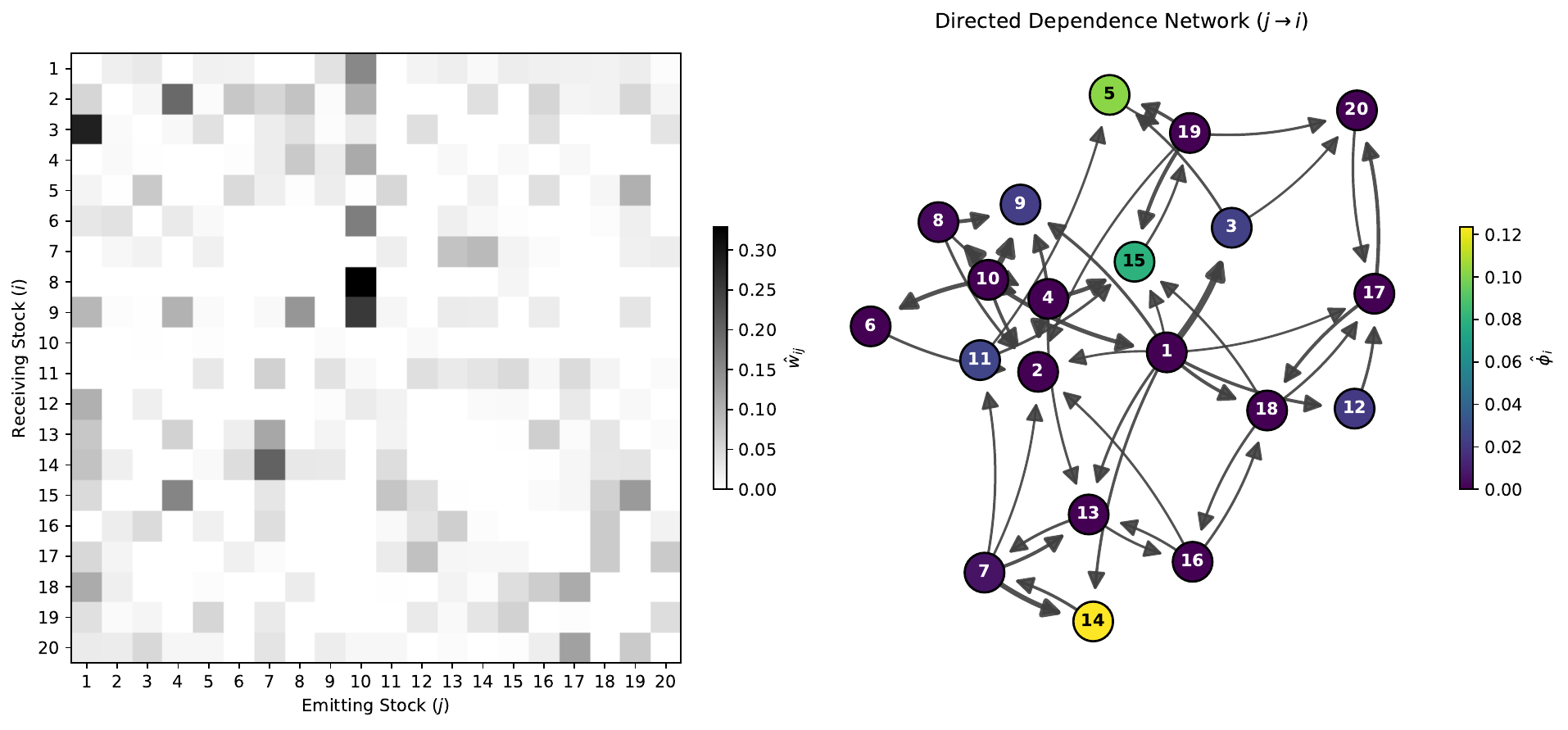}
    \caption{Estimated spatiotemporal dependence among twenty UK-listed stocks. The left panel shows the complete estimated dependence matrix $\widehat{\mathbf{W}}$, where $\widehat{w}_{ij}$ represents dependence from firm $j$ to firm $i$. The right panel displays the 50 strongest directed links, with edge width indicating dependence magnitude and node colour representing the temporal autoregressive coefficient $\widehat{\phi}(\boldsymbol{s}_i)$.}
    \label{fig:spatiotemporal_network}
\end{figure}
Figure \ref{fig:spatiotemporal_network} presents the estimated volatility dependence structure. The left panel shows the complete estimated matrix $\widehat{\mathbf{W}}$, where $\widehat{w}_{ij}$ measures the contemporaneous contribution of firm $j$ to the conditional log-volatility of firm $i$. Larger values indicate stronger dependence relative to the other estimated entries. The right panel displays the 50 strongest directed links, ranked by $\lvert\widehat{w}_{ij}\rvert$. These links represent approximately $23.1\%$ of the 216 estimated non-zero edges and are shown to preserve the visibility of edge directions and node labels. This restriction applies only to the visualisation; all 216 estimated links are retained in the model and subsequent analyses. Edge width represents the magnitude of the estimated dependence, while node colour represents the firm-specific temporal autoregressive coefficient $\widehat{\phi}(\boldsymbol{s}_i)$.

To quantify the sparsity of the estimated network, its density is defined as
\begin{equation}
\operatorname{Density}=\frac{\#\{(i,j):\widehat{w}_{ij}\neq 0,\;i\neq j\}}{n(n-1)}.\label{eq:network_density}
\end{equation}
The estimated network contains 216 of the 380 possible directed links, corresponding to a density of $0.568$. Thus, the penalisation removes approximately $43.2\%$ of the potential links. Although the network remains moderately dense, the one-standard-error rule produces a more parsimonious structure than both the fully connected specification and the minimum-error model. The retained relationships exhibit considerable heterogeneity in magnitude and are predominantly cross-sectoral, with $78.7\%$ of the estimated dependence occurring between firms in different sectors and $21.3\%$ occurring within sectors. The estimated structure is also directional because, in general, $\widehat{w}*{ij}\neq\widehat{w}*{ji}$. Glencore (GLEN.L; index 10) is associated with several of the strongest individual links, including outgoing links to Rio Tinto (RIO.L; index 8) and Anglo American (AAL.L; index 9). The model therefore distinguishes firms that primarily transmit volatility dependence from those that primarily receive it. These relationships represent estimated conditional dependence and should not be interpreted as evidence of causal transmission.

\subsection{Network Centrality}

\begin{figure}[!t]
    \centering
    \includegraphics[width=\linewidth]{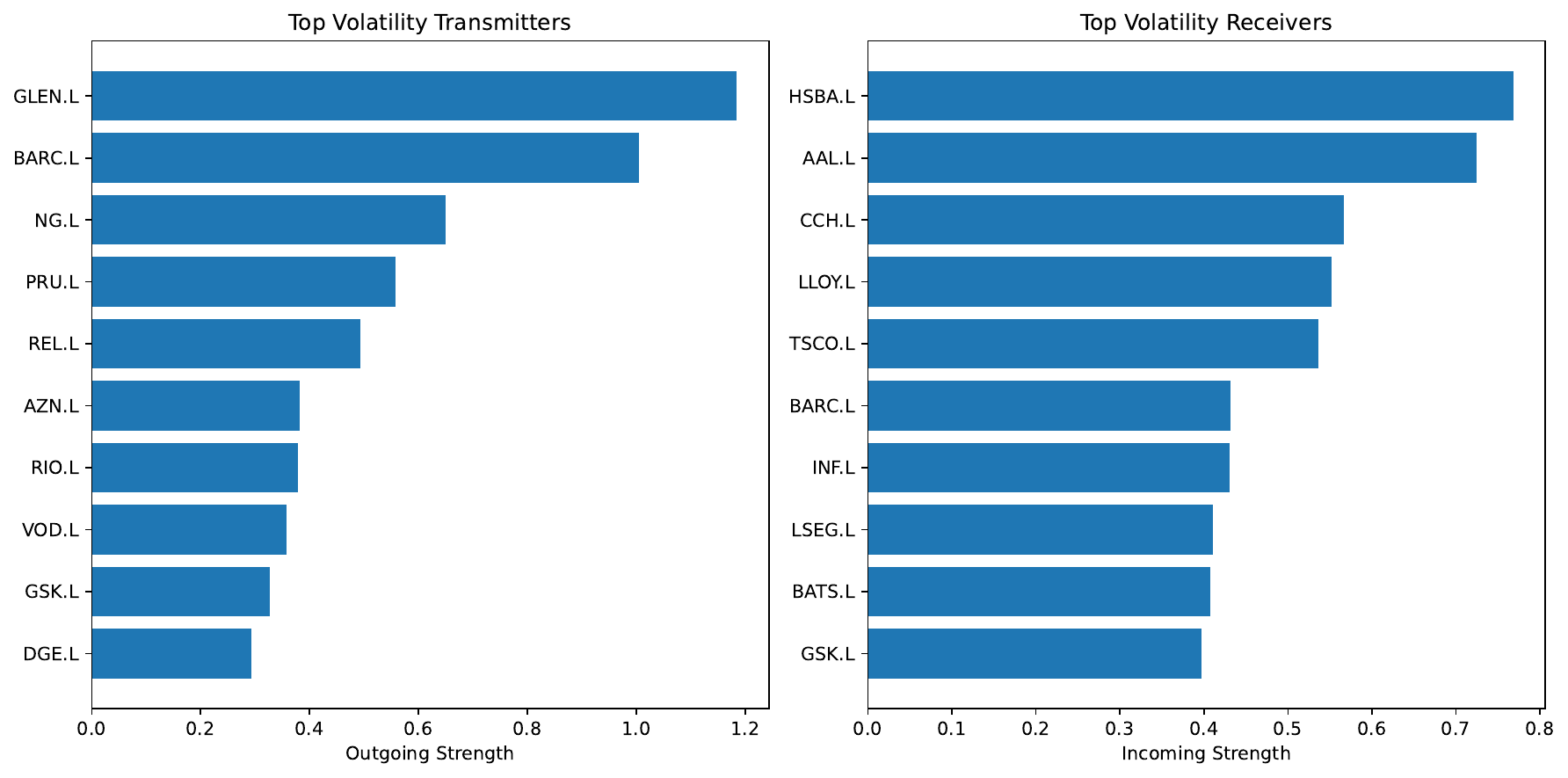}
    \caption{Estimated outgoing and incoming volatility-dependence strengths. Larger values identify firms with stronger transmitting and receiving roles, respectively.}
    \label{fig:network_centrality}
\end{figure}
To characterise the overall role of each firm, outgoing and incoming dependence strengths are defined as
\begin{equation}
C_i^{\mathrm{out}}=\sum_{j\neq i}\widehat{w}_{ji},\qquad C_i^{\mathrm{in}}=\sum_{j\neq i}\widehat{w}_{ij},
\label{eq:centrality_measures}
\end{equation}

where $C_i^{\mathrm{out}}$ measures the total dependence contributed by firm $i$ to other firms, while $C_i^{\mathrm{in}}$ measures its total exposure to dependence originating from other firms.
\begin{table}[!t]
\centering
\caption{Firm-level dependence strengths and temporal autoregressive coefficients.}
\label{tab:centrality}
\small
\resizebox{\textwidth}{!}{
\begin{tabular}{lllrrr}
\toprule
\textbf{Company} & \textbf{Ticker} & \textbf{Sector} & \textbf{Outgoing} & \textbf{Incoming} & $\widehat{\phi}(s_i)$ \\
\midrule
Glencore                    & GLEN.L & Energy/Resources   & 1.185 & 0.019 & $<0.0001$ \\
Barclays                    & BARC.L & Banking/Financials & 1.005 & 0.432 & $<0.0001$ \\
National Grid               & NG.L   & Energy/Resources   & 0.650 & 0.303 & 0.0061 \\
Prudential                  & PRU.L  & Banking/Financials & 0.557 & 0.272 & $<0.0001$ \\
RELX                        & REL.L  & Telecom \& Pharma  & 0.493 & 0.318 & $<0.0001$ \\
AstraZeneca                 & AZN.L  & Telecom \& Pharma  & 0.382 & 0.368 & $<0.0001$ \\
Rio Tinto                   & RIO.L  & Energy/Resources   & 0.379 & 0.342 & 0.0028 \\
Vodafone Group              & VOD.L  & Telecom \& Pharma  & 0.357 & 0.329 & $<0.0001$ \\
GlaxoSmithKline             & GSK.L  & Telecom \& Pharma  & 0.327 & 0.398 & $<0.0001$ \\
Diageo                      & DGE.L  & Consumer Goods     & 0.293 & 0.223 & 0.0207 \\
British American Tobacco    & BATS.L & Consumer Goods     & 0.290 & 0.408 & $<0.0001$ \\
Tesco                       & TSCO.L & Consumer Goods     & 0.288 & 0.536 & 0.1235 \\
Unilever                    & ULVR.L & Consumer Goods     & 0.286 & 0.368 & 0.0260 \\
Lloyds Banking Group        & LLOY.L & Banking/Financials & 0.267 & 0.552 & 0.0241 \\
Coca-Cola HBC               & CCH.L  & Consumer Goods     & 0.240 & 0.566 & 0.0792 \\
London Stock Exchange Group & LSEG.L & Banking/Financials & 0.230 & 0.411 & 0.1021 \\
Shell                       & SHEL.L & Energy/Resources   & 0.228 & 0.332 & $<0.0001$ \\
Informa                     & INF.L  & Telecom \& Pharma  & 0.220 & 0.431 & $<0.0001$ \\
Anglo American              & AAL.L  & Energy/Resources   & 0.214 & 0.724 & 0.0223 \\
HSBC Holdings               & HSBA.L & Banking/Financials & 0.207 & 0.768 & $<0.0001$ \\
\bottomrule
\end{tabular}
}
\end{table}

Figure \ref{fig:network_centrality} and Table \ref{tab:centrality} reveal marked differences between firms’ transmitting and receiving roles. Glencore (GLEN.L) has the largest outgoing strength (1.185), followed by Barclays (BARC.L; 1.005), National Grid (NG.L; 0.650), Prudential (PRU.L; 0.557), and RELX (REL.L; 0.493). In contrast, HSBC Holdings (HSBA.L) has the largest incoming strength (0.768), followed by Anglo American (AAL.L; 0.724), Coca-Cola HBC (CCH.L; 0.566), Lloyds Banking Group (LLOY.L; 0.552), and Tesco (TSCO.L; 0.536). The contrasting rankings demonstrate the directional and asymmetric nature of the estimated network. For example, Glencore is the strongest transmitter but has the smallest incoming strength (0.019), whereas HSBC has the greatest incoming strength despite having the smallest outgoing strength (0.207). Thus, firms that strongly transmit volatility dependence are not necessarily those most exposed to spillovers originating from other firms.

\subsection{Sectoral Dependence}

To examine the network at the sector level, the estimated weights are aggregated as
\begin{equation}
S_{ab}=\sum_{i\in G_a}\sum_{\substack{j\in G_b\\j\neq i}}\widehat{w}_{ij},\label{eq:sectoral-spillover}
\end{equation}
where $G_a$ and $G_b$ denote the sets of firms in sectors $a$ and $b$, respectively. Consequently, $S_{ab}$ measures total estimated dependence from sector $b$ to sector $a$.

\begin{table}[!t]
\centering
\caption{Sector-level volatility dependence. Entry $(a,b)$ represents the total dependence transmitted from originating sector $b$ to receiving sector $a$.}
\label{tab:sector_spillovers}
\small
\begin{tabular}{lrrrr}
\toprule
\textbf{Receiving sector} & \textbf{Banking} & \textbf{Energy} & \textbf{Consumer} & \textbf{Telecom/Pharma} \\
\midrule
Banking/Financials & 0.7721 & 0.9434 & 0.2416 & 0.4774 \\
Energy/Resources  & 0.3657 & 0.8975 & 0.3252 & 0.1325 \\
Consumer Goods    & 0.5963 & 0.6245 & 0.3659 & 0.5154 \\
Telecom/Pharma    & 0.5322 & 0.1914 & 0.4650 & 0.6541 \\
\bottomrule
\end{tabular}
\end{table}

Table \ref{tab:sector_spillovers} shows that volatility dependence extends beyond sectoral boundaries. The largest aggregate cross-sector relationship runs from Energy/Resources to Banking/Financials ($0.9434$), while Energy/Resources also exhibits the strongest within-sector dependence ($0.8975$). Other notable cross-sector relationships include transmission from Energy/Resources to Consumer Goods ($0.6245$) and from Banking/Financials to Consumer Goods ($0.5963$). These sizeable off-diagonal entries indicate that sector membership alone does not adequately represent the observed dependence structure, supporting the data-driven estimation of $\mathbf{W}$.

\subsection{Temporal Volatility Persistence}

The model estimates firm-specific temporal autoregressive coefficients $\widehat{\phi}(\boldsymbol{s}_i)$, which measure own-lag persistence after accounting for contemporaneous dependence through $\widehat{\mathbf{W}}$. It also estimates unit-specific intercepts $\widehat{\mu}(\boldsymbol{s}_i)$, which capture heterogeneity in baseline log conditional variance.
\begin{figure}[!t]
    \centering
    \includegraphics[width=\linewidth]{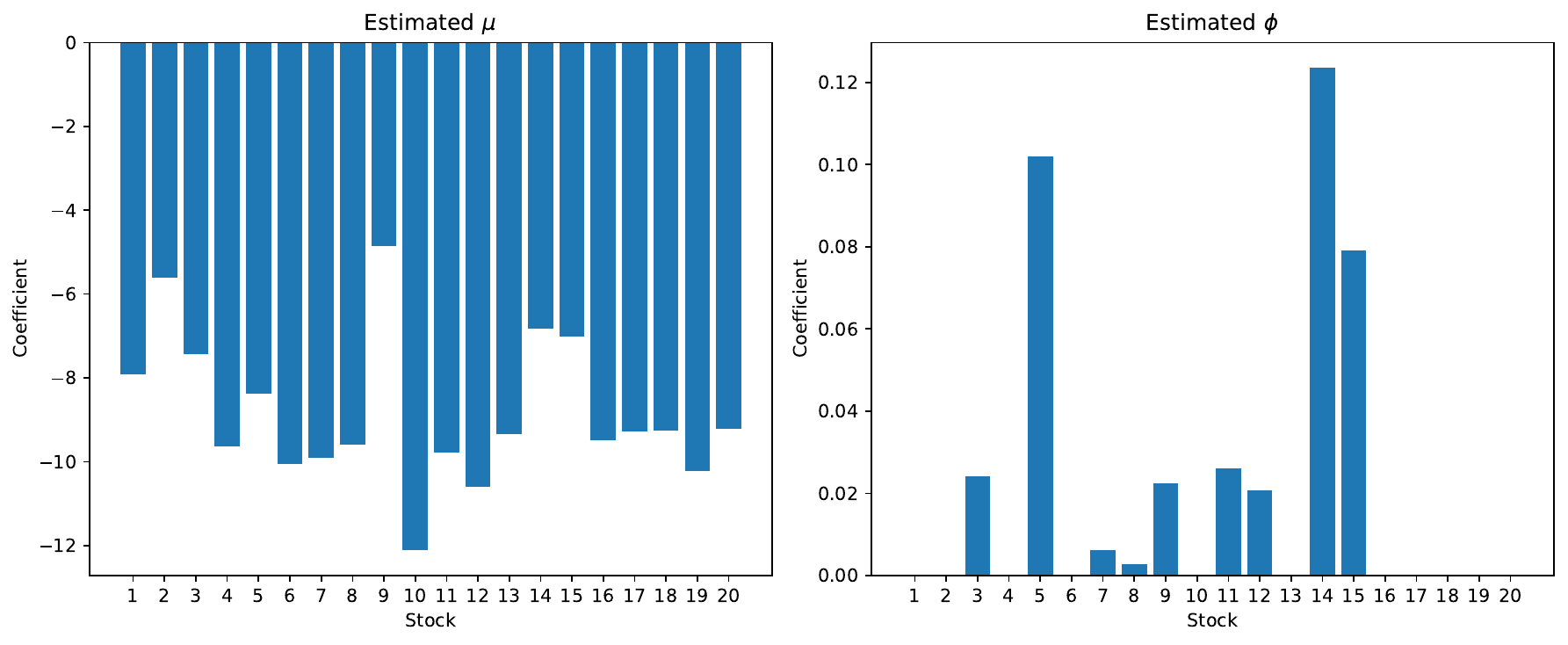}
    \caption{Estimated unit-specific intercepts $\widehat{\mu}(\boldsymbol{s}_i)$ (left) and temporal autoregressive coefficients $\widehat{\phi}(\boldsymbol{s}_i)$ (right) for the twenty firms.}
    \label{fig:phi_U_plot}
\end{figure}
Figure \ref{fig:phi_U_plot} persistence is concentrated in a subset of firms. Tesco (TSCO.L; index 14) has the largest temporal autoregressive coefficient ($0.1235$), followed by London Stock Exchange Group (LSEG.L; index 5; $0.1021$) and Coca-Cola HBC (CCH.L; index 15; $0.0792$). Nine firms retain non-zero temporal coefficients, whereas the remaining eleven estimates are effectively zero. Thus, after accounting for contemporaneous network dependence, own-lag volatility contributes meaningfully to the conditional variance dynamics of fewer than half of the firms. The estimated intercepts also vary across firms, indicating heterogeneity in baseline log conditional variance.

\subsection{Comparison with Predefined Dependence Structures}\label{sec:comparison_W}

The estimated $\widehat{\mathbf{W}}$ was compared with predefined Euclidean-distance, correlation-based, autoregressive-similarity, and sector-based matrices, including sparse $k$-nearest-neighbour variants. All models were estimated using the same data and evaluated using likelihood-based and out-of-sample predictive criteria. For the penalised model, AIC and BIC were computed using the effective parameter count $p_{\mathrm{eff}}=\|\widehat{\boldsymbol{\beta}}\|_0+n+\|\widehat{\boldsymbol{\phi}}\|_0+\|\widehat{\mathbf{W}}\|_0$, where $\|\cdot\|_0$ denotes the number of non-zero estimated elements and the $n$ unit-specific intercepts are included as unpenalised parameters. The effective sample size was taken as $N=n(T_{\mathrm{train}}-1)$.

\begin{figure}[!h]
    \centering
    \includegraphics[width=0.85\textwidth]{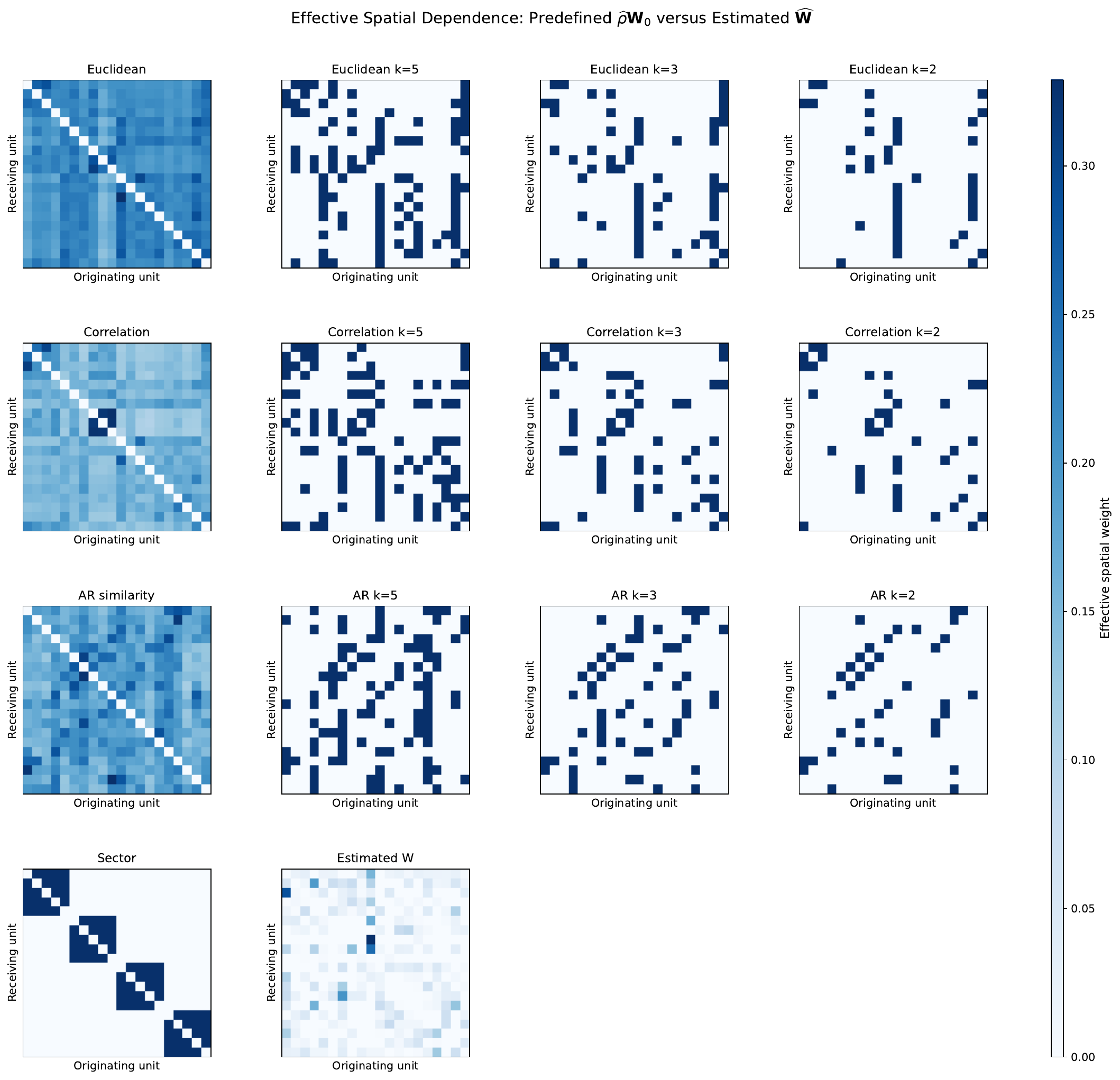}
    \caption{Comparison of effective contemporaneous dependence structures under the predefined and estimated spatial weighting schemes. For each predefined specification, the effective dependence matrix is $\widehat{\rho}\mathbf{W}_0$, where $\widehat{\rho}$ is the estimated spatial autoregressive coefficient and $\mathbf{W}_0$ is the corresponding Euclidean-distance, return-correlation, autoregressive-similarity, or sector-based weighting matrix. The $k$-nearest-neighbour variants retain $k\in\{2,3,5\}$ connections for each firm. The final panel shows the data-driven matrix $\widehat{\mathbf{W}}$ estimated by the penalised model. Columns represent originating firms and rows represent receiving firms, while darker cells indicate stronger effective dependence.}
    \label{fig:effective_spatial_dependence}
\end{figure}

Table \ref{tab:W_comparison} shows that the estimated $\mathbf{W}$ model provides the strongest out-of-sample predictive performance, with a test RMSE of $2.6716$ and MAE of $1.8051$. The closest predefined alternative, Euclidean $k=5$, yields an RMSE of $2.9334$ and MAE of $2.2645$, corresponding to reductions of approximately $8.9\%$ and $20.2\%$, respectively. The estimated $\mathbf{W}$ model also achieves the highest log-likelihood, indicating better in-sample fit. However, the correlation-based model has the lowest AIC and BIC because it is more parsimonious. Thus, while the estimated $\mathbf{W}$ model is more complex, it provides substantially better out-of-sample predictions. Pairwise tests of out-of-sample prediction loss further show significantly lower loss for the estimated $\mathbf{W}$ model relative to every predefined specification ($p<0.001$ after adjustment for multiple comparisons). These results provide empirical evidence that learning the contemporaneous dependence structure can yield meaningful predictive gains when the underlying network is unknown. Figure \ref{fig:effective_spatial_dependence} compares the effective dependence structures implied by the predefined weighting schemes, $\widehat{\rho}\mathbf{W}_0$, with the data-driven estimate $\widehat{\mathbf{W}}$.

\begin{table}[!t]
\centering
\caption{Predictive comparison of the estimated dependence structure and selected predefined specifications.}
\label{tab:W_comparison}
\small
\begin{tabular}{lrrrrr}
\toprule
Model & LogLik & AIC & BIC & Test RMSE & Test MAE \\
\midrule
Estimated $\mathbf{W}$ & \textbf{-19950.83} & 40393.66 & 42111.90 & \textbf{2.6716	} & \textbf{1.8051} \\
Euclidean ($k=5$) & -20195.07 & 40474.13 & 40767.49 & 2.9334 & 2.2645 \\
Euclidean & -20177.60 & 40439.21 & 40732.56 & 2.9338 & 2.2645 \\
Correlation & -20154.52 & \textbf{40393.03} & \textbf{40686.39} & 2.9339 & 2.2646 \\
AR similarity & -20194.97 & 40473.94 & 40767.30 & 2.9346 & 2.2654 \\
Sector & -20264.30 & 40612.59 & 40905.95 & 2.9363 & 2.2678 \\
\bottomrule
\end{tabular}
\end{table}
\section{Conclusion} \label{sec:conclusion}

This paper developed a penalised framework for learning an unknown contemporaneous dependence structure in a spatiotemporal conditional-variance model. Rather than specifying the spatial weights matrix $\mathbf{W}$ a priori, the proposed approach jointly estimates its elements with temporal persistence, covariate effects, and unit-specific baseline components using an $\ell_1$-penalised quasi-maximum likelihood criterion. Monte Carlo experiments demonstrate good finite-sample performance, with estimation accuracy generally improving as the temporal sample size increases, although computational cost increases with the cross-sectional dimension.

The empirical application to twenty UK-listed firms demonstrates the practical value of learning the dependence structure directly from the data. The estimated network reveals heterogeneous and directional volatility dependence, including both within-sector and cross-sector relationships. Importantly, the estimated-$\mathbf{W}$ model achieves substantially lower out-of-sample prediction errors than all predefined dependence structures considered, with these improvements statistically significant in the pairwise comparisons. Although the correlation-based specification yields lower AIC and BIC because of its greater parsimony, the results indicate that the additional flexibility of learning $\mathbf{W}$ provides meaningful predictive gains.

The findings support estimating the contemporaneous dependence structure when the underlying network is unknown rather than imposing it a priori. The current framework assumes a time-invariant, non-negative dependence matrix and becomes computationally demanding as the network dimension increases. Future work could therefore consider time-varying dependence structures, alternative regularisation methods, scalable optimisation strategies, and the estimator's theoretical properties in high-dimensional settings.

\section*{Data Availability Statement}

The financial data analysed in this study were obtained from Yahoo Finance and are publicly accessible. The simulated data and computer code required to reproduce the analyses and results can be requested from the corresponding author. 





\section*{Conflict of Interest}

The authors declare no conflict of interest.

\bibliographystyle{abbrvnat}
\bibliography{references}

\begin{thebibliography}{28}
\providecommand{\natexlab}[1]{#1}
\providecommand{\url}[1]{\texttt{#1}}
\expandafter\ifx\csname urlstyle\endcsname\relax
  \providecommand{\doi}[1]{doi: #1}\else
  \providecommand{\doi}{doi: \begingroup \urlstyle{rm}\Url}\fi

\bibitem[Ahrens and Bhattacharjee(2015)]{ahrens2015two}
A.~Ahrens and A.~Bhattacharjee.
\newblock Two-step lasso estimation of the spatial weights matrix.
\newblock \emph{Econometrics}, 3\penalty0 (1):\penalty0 128--155, 2015.

\bibitem[Bauwens et~al.(2006)Bauwens, Laurent, and Rombouts]{bauwens2006multivariate}
L.~Bauwens, S.~Laurent, and J.~V. Rombouts.
\newblock Multivariate {GARCH} models: a survey.
\newblock \emph{Journal of {A}pplied {E}conometrics}, 21\penalty0 (1):\penalty0 79--109, 2006.

\bibitem[Bhattacharjee and Jensen-Butler(2006)]{bhattacharjee2006estimation}
A.~Bhattacharjee and C.~Jensen-Butler.
\newblock Estimation of spatial weights matrix, with an application to diffusion in housing demand.
\newblock \emph{Centre for Research into Industry, Enterprise, Finance, and the Firm Discussion Paper}, 519, 2006.

\bibitem[Bhattacharjee and Jensen-Butler(2013)]{bhattacharjee2013estimation}
A.~Bhattacharjee and C.~Jensen-Butler.
\newblock Estimation of the spatial weights matrix under structural constraints.
\newblock \emph{Regional Science and Urban Economics}, 43\penalty0 (4):\penalty0 617--634, 2013.

\bibitem[Bollerslev(1986)]{bollerslev1986generalized}
T.~Bollerslev.
\newblock Generalized autoregressive conditional heteroskedasticity.
\newblock \emph{Journal of {E}conometrics}, 31\penalty0 (3):\penalty0 307--327, 1986.

\bibitem[Cressie and Wikle(2011)]{cressie2011statistics}
N.~Cressie and C.~K. Wikle.
\newblock \emph{Statistics for Spatio-Temporal Data}.
\newblock John Wiley \& Sons, 2011.

\bibitem[Diebold and Yilmaz(2009)]{diebold2009measuring}
F.~X. Diebold and K.~Yilmaz.
\newblock Measuring financial asset return and volatility spillovers, with application to global equity markets.
\newblock \emph{The Economic Journal}, 119\penalty0 (534):\penalty0 158--171, 2009.

\bibitem[Diebold and Yilmaz(2012)]{diebold2012better}
F.~X. Diebold and K.~Yilmaz.
\newblock Better to give than to receive: Predictive directional measurement of volatility spillovers.
\newblock \emph{International {J}ournal of {F}orecasting}, 28\penalty0 (1):\penalty0 57--66, 2012.

\bibitem[Diebold and Y{\i}lmaz(2014)]{diebold2014network}
F.~X. Diebold and K.~Y{\i}lmaz.
\newblock On the network topology of variance decompositions: Measuring the connectedness of financial firms.
\newblock \emph{Journal of {E}conometrics}, 182\penalty0 (1):\penalty0 119--134, 2014.

\bibitem[Engle(2002)]{engle2002dynamic}
R.~Engle.
\newblock Dynamic conditional correlation: A simple class of multivariate generalized autoregressive conditional heteroskedasticity models.
\newblock \emph{Journal of {B}usiness \& {E}conomic {S}tatistics}, 20\penalty0 (3):\penalty0 339--350, 2002.

\bibitem[Engle(1982)]{engle1982autoregressive}
R.~F. Engle.
\newblock Autoregressive conditional heteroscedasticity with estimates of the variance of {U}nited {K}ingdom inflation.
\newblock \emph{Econometrica: Journal of the {E}conometric {S}ociety}, pages 987--1007, 1982.

\bibitem[Engle and Kroner(1995)]{engle1995multivariate}
R.~F. Engle and K.~F. Kroner.
\newblock Multivariate simultaneous generalized {ARCH}.
\newblock \emph{Econometric Theory}, 11\penalty0 (1):\penalty0 122--150, 1995.

\bibitem[Forbes and Rigobon(2002)]{forbes2002no}
K.~J. Forbes and R.~Rigobon.
\newblock No contagion, only interdependence: measuring stock market comovements.
\newblock \emph{The Journal of Finance}, 57\penalty0 (5):\penalty0 2223--2261, 2002.

\bibitem[Johnson et~al.(1994)Johnson, Kotz, and Balakrishnan]{johnson1994continuous}
N.~L. Johnson, S.~Kotz, and N.~Balakrishnan.
\newblock \emph{Continuous Univariate Distributions, volume 1}, volume~1.
\newblock John Wiley \& Sons, 1994.

\bibitem[Lam and Souza(2020)]{lam2020estimation}
C.~Lam and P.~C. Souza.
\newblock Estimation and selection of spatial weight matrix in a spatial lag model.
\newblock \emph{Journal of {B}usiness \& {E}conomic {S}tatistics}, 38\penalty0 (3):\penalty0 693--710, 2020.

\bibitem[LeSage and Pace(2009)]{lesage2009introduction}
J.~LeSage and R.~K. Pace.
\newblock \emph{Introduction to Spatial Econometrics}.
\newblock Chapman and Hall/CRC, 2009.

\bibitem[Meinshausen and B{\"u}hlmann(2006)]{meinshausen2006high}
N.~Meinshausen and P.~B{\"u}hlmann.
\newblock {High-dimensional graphs and variable selection with the Lasso}.
\newblock \emph{The Annals of Statistics}, 34\penalty0 (3):\penalty0 1436 -- 1462, 2006.
\newblock \doi{10.1214/009053606000000281}.
\newblock URL \url{https://doi.org/10.1214/009053606000000281}.

\bibitem[Merk and Otto(2022)]{merk2022estimation}
M.~S. Merk and P.~Otto.
\newblock Estimation of the spatial weighting matrix for regular lattice data—an adaptive {L}asso approach with cross-sectional resampling.
\newblock \emph{Environmetrics}, 33\penalty0 (1):\penalty0 e2705, 2022.

\bibitem[Otto(2024)]{otto2024multivariate}
P.~Otto.
\newblock A multivariate spatial and spatiotemporal {ARCH} model.
\newblock \emph{Spatial Statistics}, 60:\penalty0 100823, 2024.

\bibitem[Otto and Steinert(2023)]{otto2023estimation}
P.~Otto and R.~Steinert.
\newblock Estimation of the spatial weighting matrix for spatiotemporal data under the presence of structural breaks.
\newblock \emph{Journal of Computational and Graphical Statistics}, 32\penalty0 (2):\penalty0 696--711, 2023.

\bibitem[Otto et~al.(2018)Otto, Schmid, and Garthoff]{otto2018generalised}
P.~Otto, W.~Schmid, and R.~Garthoff.
\newblock Generalised spatial and spatiotemporal autoregressive conditional heteroscedasticity.
\newblock \emph{Spatial Statistics}, 26:\penalty0 125--145, 2018.

\bibitem[Otto et~al.(2024)Otto, Do{\u{g}}an, and Ta{\c{s}}p{\i}nar]{otto2024dynamic}
P.~Otto, O.~Do{\u{g}}an, and S.~Ta{\c{s}}p{\i}nar.
\newblock Dynamic spatiotemporal {ARCH} models.
\newblock \emph{Spatial {E}conomic Analysis}, 19\penalty0 (2):\penalty0 250--271, 2024.

\bibitem[Otto et~al.(2025)Otto, Do{\u{g}}an, Ta{\c{s}}p{\i}nar, Schmid, and Bera]{otto2025spatial}
P.~Otto, O.~Do{\u{g}}an, S.~Ta{\c{s}}p{\i}nar, W.~Schmid, and A.~K. Bera.
\newblock Spatial and spatiotemporal volatility models: A review.
\newblock \emph{Journal of Economic Surveys}, 39\penalty0 (3):\penalty0 1037--1091, 2025.

\bibitem[Sato and Matsuda(2017)]{sato2017spatial}
T.~Sato and Y.~Matsuda.
\newblock Spatial autoregressive conditional heteroskedasticity models.
\newblock \emph{Journal of the Japan Statistical Society}, 47\penalty0 (2):\penalty0 221--236, 2017.

\bibitem[Sato and Matsuda(2021)]{sato2021spatial}
T.~Sato and Y.~Matsuda.
\newblock Spatial extension of generalized autoregressive conditional heteroskedasticity models.
\newblock \emph{Spatial {E}conomic {A}nalysis}, 16\penalty0 (2):\penalty0 148--160, 2021.
\newblock \doi{10.1080/17421772.2020.1742929}.

\bibitem[Tibshirani(1996)]{tibshirani1996regression}
R.~Tibshirani.
\newblock Regression shrinkage and selection via the {L}asso.
\newblock \emph{Journal of the Royal Statistical Society: Series B (Methodological)}, 58\penalty0 (1):\penalty0 267--288, 1996.

\bibitem[Van~der Weide(2002)]{van2002go}
R.~Van~der Weide.
\newblock Go-{GARCH}: a multivariate generalized orthogonal {GARCH} model.
\newblock \emph{Journal of {A}pplied {E}conometrics}, 17\penalty0 (5):\penalty0 549--564, 2002.

\bibitem[Zhang and Yu(2018)]{zhang2018spatial}
X.~Zhang and J.~Yu.
\newblock Spatial weights matrix selection and model averaging for spatial autoregressive models.
\newblock \emph{Journal of {E}conometrics}, 203\penalty0 (1):\penalty0 1--18, 2018.

\end{thebibliography}

\end{document}